\documentstyle[prd,aps]{revtex}
\begin{document}
\draft
\title{THE STRING MODEL OF GRAVITY}
\author{Miroslav Pardy}
\address{Department of Theoretical Physics and Astrophysics \\
Masaryk University \\
Kotl\'{a}\v{r}sk\'{a} 2, 611 37 Brno, Czech Republic\\
email:pamir@elanor.sci.muni.cs}
\date{\today}
\maketitle

\begin{abstract}
The string model of gravitational force is proposed where the
string forms the mediation of the gravitational interaction between
two gravitating bodies. It reproduces the
Newtonian results in the first-order approximation and it predicts in the
higher-oder approximations the existence of oscillations of the
gravitational field between two massive bodies. It can be easily generalized
to the two-body interaction in particle physics.\\
\end{abstract}

\section{Introduction}
\hspace{3ex}
In history of exact sciences there exist some problems which were formulated
some centuries ago and solved only in this century. In mathematics it is for
instance the Fermat theorem which was resolved at the recent time. In
physics there is
the problem of action-at-a-distance which was for the first time formulated by
Newton in his "Principia Mathematica" [1] and never solved by him.

While the Fermat theorem is extremaly difficult to prove,
the action-at-a-distance problem can be
solved because it is enough to use approximation.

Instead of resolution of this problem
Newton suggested the phenomenological theory of the gravitational force
where there exists no answer concerning the dynamics or the mechanism of
action-at-a-distance. Newton himself was awared that it necessary exists
some mediation of interaction between two bodies at the
different points in space because he has written [1]:
"It is inconceivable, that inanimate brute matter,
should without the mediation of something else which is not material, operate
upon and affect other matter without mutual contact .. ". In other words,
the crucial notion in the Newton speculation is the mediation between two
bodies.

By analogy with the mechanical situation we will
suppose the model where the atractive force between two
bodies is transmitted as tension in the fictitious string connecting
the one body with the another one. Then, the theoretical problem is
to show that such model works and gives not only the old results but
new results which cannot be derived from the original Newton law.

We will consider the string, the
left end of which is fixed at the beginning of the coordinate system
and mass $m$ is fixed
on the right end of the string. The motion of
the system string  and the body with mass $m$ is the
fundamental problem of the
equations of the mathematical physics in case that the tension is linearly
dependent on elongation [2]. We will show that
it is possible to represent the Newton gravitational law
by the string with the nonlinear tension in the string. Because of the strong
nonlinearity of the problem
the motion of the string and the body can be solved only approximately.
In the following text we will give the aproximative version of the
Kepler problem and then we get the string solution of this problem.

\section{The Kepler problem}
\hspace{3ex}

Let us consider two bodies 1 and 2 with masses $M$ and $m$, where $M \gg m$.
The body 1 is supposed to be fixed at the origin of the coordinate system and
the body 2 is for the simplicity moving in the interval

$$(R-\delta, R+\delta),
\eqno(1)$$

\noindent
where $\delta \ll R$, which corresponds to the motion of planets of our Sun
system. The Newton law

$$F = -\kappa \frac {M\*m}{r^2},
\eqno(2)$$
can be obviously expressed in the interval (1) approximatelly  as

$$F \approx a\*\eta+b; \quad (-\delta,\delta)\owns\eta,
\eqno(3)$$
where

$$a = \frac {2\kappa Mm}{R^3}, \quad b = -\frac {\kappa Mm}{R^2}.
\eqno(4)$$

The motion of body 2 in the gravitational potential of body 1 is
described by equation [3]

$$m\stackrel{..}{r} = -\kappa\frac {Mm}{r^2} + \frac {J^2}{mr^3},
\eqno(5)$$
where $J$ is the angular momentum of body 2. In the inerval $(-\delta,\delta)$
we can write

$$r(t) = R + \eta(t)\eqno(6)$$
and using approximation

$$\frac {1}{(R+\eta)^2} \approx \frac {1}{R^2}(1-\frac {2\eta}{R}),
\quad \frac {1}{(R+\eta)^3} \approx \frac {1}{R^3}(1-\frac {3\eta}{R}),
\eqno(7)$$
we get after insertion of eq. (6) into eq. (5):

$$\stackrel{..}{\eta} + \omega^2\eta = \lambda, \eqno(8)$$
where

$$\omega^2 = \frac {3J^2}{m^2\*R^4} - \frac{2\kappa\*M}{R^3}\eqno(9)$$

$$\lambda = \frac {J^2}{m^2\*R^3} - \frac {\kappa\*M}{R^2}. \eqno(10)$$

For the circle motion we have $J= m\omega\*R^2, r = R$ and from eqs.
(9) and (10) it follows:

$$\omega = R^{-3/2}\*(\kappa\*M)^{1/2}; \quad \lambda = 0. \eqno(11)$$

It is easy to see that the solution of eq. (8) is of the form:

$$\eta(t) = \Lambda\*\cos(\omega t + \vartheta) +\frac {\lambda}{\omega^2},
\eqno(12)$$
where $\Lambda$ and $\vartheta$ are constants involving the initial
conditions of motion of the body 2.

So far we have supposed no dynamics of mediation of the interaction
between body 1 and 2. However, only the model involving the mechanism
of mediation of interaction can describe logically consistent reality
and explain the Newton puzzle. Let us try to elaborate the consistent
and realistic model
which describes the mechanism of mediation.

\section{The string mediation of interaction}
\hspace{3ex}

In this section we will solve the motion of a body  2 at the end of the string
on the assumption that the tension in the string is nonlinear and it
generates the Newton law in the statical regime. We will give the
rigorous mathematical formulation of the problem. While for the Hook tension
the problem has solution by the Fourier method, in case of the nonlinear
tension it is not possible to use this method.
There is
no evidence about solution of this problem in the textbooks of mathematical
physics or in the mathematical journals. So, it seems,
we solve this problem for the first time.

Let be given the string, the left end of which is fixed at beginning
and the right end is at point $l$ at the state of equilibrium. The
deflection of the string element $dl$ at point $x$ and time $t$ let
be $u(x,t)$ where $x\in(0,l)$ and

$$\eta(t) = u(l,t), \quad \eta(0) = u(l,0). \eqno(13)$$

Then, the motion of body 2 is described by the motion of the right
end-point of the string, when the left point is constantly fixed at
the origin.

The differential equation of motion of string elements can be derived
by the following way: We suppose  that the force acting on
the element $dl$ of the string is given by the law:

$$T(x,t) = -ES\left(\frac {\partial \*u}{\partial x}\right)^{-2}, \eqno(14)$$
where $E$ is the modulus of elasticity, $S$ is the cross section of
the string. We easily derive that

$$T(x+dx)-T(x) = 2ES \left(u_{x}\right)^{-3}u_{xx}dx. \eqno(15)$$

The mass $dm$  of the element $dl$ is $\varrho ESdx$,
where $\varrho$ is the mass
density of the string matter and the dynamical equilibrium gives

$$\varrho\*Sdx u_{tt} = 2ESu_{xx}\left(u_{x}\right)^{-3}dx. \eqno(16)$$

Putting

$$\varrho = \varrho_{0}\frac {2}{\left(u_{x}\right)^{3}};\quad
\varrho_{o} = const.,\eqno(17)$$
we get

$$\frac {1}{c^2}u_{tt} - u_{xx} = 0; \quad
c  = \left(\frac {E}{\varrho_{0}}\right)^{1/2}. \eqno(18)$$

The last procedure was performed evidently in order to get the
wave equation.

Now, let us look for the correspondence between the string tension and the
Newton law. Putting $u_{tt} = 0$ we get the stationary case with the solution

$$u(x,t) = \alpha x + \beta. \eqno(19)$$

Because $u(0,t) \equiv 0$, we get $\beta=0$. Then $u_{x}(x,t) = \alpha$ is not
dependent on $x$ and according to the definition of the tension the force is
constant along the length of the string which is the same result as in the
case with the Hook law.

For sufficiently big elongation we have
$u(l)\gg l$ and the elongation at point $l$ is the distance of the right end
of the string from the origin and it means that the force acting on the right
end of the string is proportional to the minus square of the distance of the
right end of the string as in the Newton gravitational law. So we have
demonstrated that the Newton gravitational force can be simulated by the
string however with the difference that in the Newton force there is no
mediation between two bodies while in our case we have mediation caused by the
medium of the string. Now we can repeat the formulation of the problem
dscribed in the previous section in such a way that we will use the
dynamical equation (18) instead of eq. (5).
So, let us approach the solution of the problem
of the motion of body on the end of the string where the tension of the
string is defined by equation (14).

From (19) we have:

$$\alpha = \frac {u(l,t)}{l}.\eqno(20)$$

Thus,

$$T(l,t) =  -\frac {ESl^{2}}{u^{2}(l,t)} = -\kappa \frac {m M}{u^{2}(l,t)},
\eqno(21)$$
which gives the relation between the string constants and the gravitating
parameters

$$ESl^{2} =\kappa m M. \eqno(22)$$

The complete solution of eq. (18) includes the initial and boundary
conditions. The simplest nontrivial initial conditions can be chosen
with regard to the character of the problem and they are:

$$u(x,0) = \frac {R}{l} x, \quad u_{t}(x,0) = 0. \eqno(23)$$

The boundary conditions are given with respect to the dynamical
equation (5):

$$u(0,t) = 0, \quad mu_{tt}(l,t) = T(l,t) + \frac {J^{2}}{mu^{3}(l,t)}.
\eqno(24)$$

The solution of the wave equation with the strongly nonlinear boundary
conditions is evidently beyond the possibility of the present mathematical
physics. Nor the Fourier method, nor the d'Alembert one can be used in
solution of our problem. So we are forced to find only the approximation
of this problem. For this goal we write:

$$u(x,t) = \frac {R}{l}x + v(x,t), \eqno(25)$$
from which follows

$$u_{x}(x,t) = \frac {R}{l} + v_{x},\quad u(l,t) = R + v \eqno(26)$$
and we suppose that $v \ll R$. In such a way the intial conditions are:

$$v(x,0) = 0,\quad v_{t}(x,0) = 0. \eqno(27)$$

The approximative formulae are given in the following form:

$$\frac {1}{u_{x}^{2}(x,t)} \approx \frac {l^{2}}{R^{2}}
 - \frac {2v_{x}l^{3}}{R^{3}},\eqno(28)$$

$$\frac {1}{u^{3}(x,t)} \approx \frac {1}{R^{3}} - \frac {3v}{R^{4}}.
\eqno(29)$$

So, we get the new problem of mathematical physics: the wave equation

$$v_{tt} = c^{2}v_{xx}\eqno(30)$$
with the initial conditions

$$v(x,0) = 0; \quad v_{t}(x,0) = 0\eqno(31)$$
and with the boundary conditions

$$v(0,t) = 0; \quad mv_{tt}(l,t) = a + bv_{x}(l,t) + dv(l,t),\eqno(32)$$
where we have put

$$a = -\kappa\frac {Mm}{R^{2}} + \frac {J^2}{m\*R^{3}};
\quad b = \frac {2\kappa Mm}{R^{3}}l;\quad
d = -\frac {3J^{2}}{mR^{4}}.\eqno(33)$$

The equation (30) with the initial and boundary conditions (31) and (32)
represents one of the standard problems of the mathematical physics and can
be easily solved using the Laplace transform [4]:

$$\hat L u(x,t) \stackrel{d}{=}\int_{0}^{\infty}e^{-pt}
u(x,t)dt \stackrel{d}{=} u(x,p). \eqno(34)$$

Using (30) we get with $\hat L v(x,t) \stackrel {d}{=} \varphi(x,p)$:

$$\hat L v_{tt}(x,t) = p^{2}\varphi(x,p) - pv(x,0) - v_{t}(x,0) =
p^{2}\varphi(x,p)\eqno(35)$$

$$\hat L v_{xx}(x,t) =  \varphi_{xx}(x,p);\quad
\hat L a = \frac {a}{p};\quad\hat L v(0,t) =  \varphi(0,p) = 0. \eqno(36)$$

After elementary mathematical operations we get the differential
equation for $\varphi$ in the form:

$$\varphi_{xx}(x,p) - k^{2}\varphi(x,p) = 0; \quad
k = p/c. \eqno(37)$$
with the boundary condition in eq. (36).

We are looking for the the solution of eq. (37) in the form

$$\varphi(x,p) =c_{1}\cosh k x + c_{2}\sinh k x .\eqno(38)$$

We get from the boundary conditions in eq. (36) $c_{1} = 0$
and

$$c_{2} = \frac {a}{p}\;\frac {1}{(mp^{2}-d)\sinh k l -
bk\cosh k l}.\eqno(39)$$

The corresponding $\varphi(x,p)$ is of the form:

$$\varphi(x,p)= \frac {a}{p}\;\frac {\sinh k x}{(mp^{2}-d)\sinh k l -
bk\cosh k l}\eqno(40)$$

The corresponding function $v(x,t)$  follows from the theory of the Laplace
transform as the mathematical formula:

$$v(x,t) = \frac {1}{2\pi i}\oint e^{pt}\varphi(x,p) dp =
\sum_{p=p_{n}}res\;e^{pt}\varphi(x,p) = $$

$$\sum _{p=p_{n}}res\;e^{pt}\frac {a}{p}\;\frac {\sinh k x}
{(mp^{2}-d)\sinh k l - bk\cosh k l},\eqno(41)$$
where $p_{n}$ are poles of the function $\varphi(x,p)$ and they are evidently
given by equation

$$\left[(mp^{2}-d)\sinh k l - bk\cosh k l\right] = 0,
\eqno(42)$$
which is equivalent with $k \rightarrow ik$ to

$$\tan kl = \frac {-bk}{mc^{2}k^{2}+d}.
\eqno(43)$$

In case of $k \ll 1$ we have two solutions: $p_{0} = 0$ and

$$p_{1/2} = \pm \left(\frac {3J^2}{m^2\*R^4} -
\frac{2\kappa\*M}{R^3}\right)^{1/2},\eqno(44)$$
which is in agreement with eq. (9) obtained by the approximation of
classical Kepler problem. Further we have got the oscillations with
frequences $p_{n}$ in the higher order approximation:

$$p_{n}\quad\rightarrow\quad \frac {n\pi }{l},\quad n \gg 1 \eqno(45)$$

At present time
it is not clear how to detect these oscillations, or, if it will be possible to
use the experimental procedures of Braginskii et al. [5] for the detection.
However, the analogous situation was in quantum
physics where the zero frequences of
vacuum was considered as meaningless till it was shown by Casimir
that they give the atractive force between two conductive plates.
Thus these frequences cannot be a'priori canceled.

\section{Discussion}

The basic heuristical idea of this article was the string realization of the
gravitational force between two bodies.

In order to realize this idea we introduced the string
of the length $l$ with the nonlinear tension which generates in the statical
situation the Newton law at the distances much
greater then is the fundamental length of the string.
We have solved this problem only
approximately because at present time the exact solution is beyond
possibilities of mathematics.

While the string with the
Hook tension has the equilibrium state, our string is not stable and the
stability of the string requires its quantisation. However, the quantization
of the string was not problem of our article. We have only shown how to solve
the Newton puzzle of gravitation.

In case that we consider the influence of the rotation of the string on the
planetary motion, we are forced write for the total energy of the
system

$$E = \frac {m}{2}\dot{r}^{2} + \frac {1}{2}(m + \frac {1}{3}\mu)r^{2}
\dot{\varphi}^{2} + U(r),\eqno(46)$$
where $\mu$ is the mass of the string and $\varphi$ is the polar coordinate.

Then, after the time-derivation of eq. (46) it follows the corresponding
equation of motion of the system and from the Lagrange equations we have:

$$\dot{\varphi} = \frac {J_{total}}{(m + \frac {1}{3}\mu)r^{2}}.
\eqno(47)$$

It means that in principle the existence of the string can be proved
experimentaly by measurement of $\dot \varphi$.

Our problem was never defined to our knowledge in the
mathematical or physical textbooks,
monographies or scientific journals. Thus, our approach is original.
It also enables to formulate the two-body problem with arbitrary nonlinear
tension in the string and it can be applied also in particle physics.

The proposed model can be also related in the modified form to the problem of
the radial motion of quarks bound by a string and used to calculate the excited
states of such system. The original solution was considered by Bardeen et al.
[6] Chodos et al. [7] and by Frampton [8]. The new analysis of such problem
was performed by Nesterenko [9]. Nobody of these authors used our approach.
So there are open way in particle physics to follow our approach.

\end{document}